\documentclass[conference]{IEEEtran}
\usepackage{cite}
\usepackage{amsmath,amssymb,amsfonts}
\usepackage{algorithmic}
\usepackage{graphicx}
\usepackage{tikz}
\usetikzlibrary{positioning}
\usepackage{textcomp}
\usepackage{xcolor}
\usepackage{braket}
\usepackage{url}
\usepackage{comment}
\usepackage[setpagesize=false,colorlinks=true,linkcolor=blue,urlcolor=red,pdftitle={Quantum Checkers},pdfauthor={Marien Raat, Luuk van den Nouweland, Matthias Müller-Brockhausen, Mike Preuss and Evert van Nieuwenburg}]{hyperref}
\hypersetup{
  colorlinks = true,
  linkcolor  = black
}
\def\BibTeX{{\rm B\kern-.05em{\sc i\kern-.025em b}\kern-.08em
    T\kern-.1667em\lower.7ex\hbox{E}\kern-.125emX}}

\usepackage{tikz}

\newcommand\copyrighttext{%
  \footnotesize \textcopyright 2025 IEEE. Personal use of this material is permitted.
  Permission from IEEE must be obtained for all other uses, in any current or future
  media, including reprinting/republishing this material for advertising or promotional
  purposes, creating new collective works, for resale or redistribution to servers or
  lists, or reuse of any copyrighted component of this work in other works.}
\newcommand\copyrightnotice{%
\begin{tikzpicture}[remember picture,overlay]
\node[anchor=south,yshift=10pt] at (current page.south) 
  {\fbox{\parbox{\dimexpr\textwidth-\fboxsep-\fboxrule\relax}{\copyrighttext}}};
\end{tikzpicture}%
}
    
\begin{document}

\title{Quantum Checkers: The Development and Analysis of a Quantum Combinatorial Game}

\author{
\IEEEauthorblockN{
Marien Raat\IEEEauthorrefmark{2}\IEEEauthorrefmark{1},
Luuk van den Nouweland\IEEEauthorrefmark{3}\IEEEauthorrefmark{1},
Matthias Müller-Brockhausen\IEEEauthorrefmark{4},
Mike Preuss\IEEEauthorrefmark{5}
and Evert van Nieuwenburg\IEEEauthorrefmark{6}}
\IEEEauthorblockA{Leiden University, Leiden, The Netherlands\\
\IEEEauthorrefmark{2}Email: \href{mailto:raat@lorentz.leidenuniv.nl}{\color{black} raat@lorentz.leidenuniv.nl} ORCID: \href{https://orcid.org/0009-0005-6521-0054}{\color{black} 0009-0005-6521-0054}
}
\IEEEauthorblockA{\IEEEauthorrefmark{3}
Email: \href{mailto:luukvdnouweland@gmail.com}{\color{black} luukvdnouweland@gmail.com}}
\IEEEauthorblockA{\IEEEauthorrefmark{4}
Email: \href{mailto:m.f.t.muller-brockhausen@liacs.leidenuniv.nl}{\color{black} m.f.t.muller-brockhausen@liacs.leidenuniv.nl} ORCID: \href{https://orcid.org/0000-0002-2107-2180}{\color{black} 0000-0002-2107-2180}
}
\IEEEauthorblockA{\IEEEauthorrefmark{5}
Email: \href{mailto:m.preuss@liacs.leidenuniv.nl}{\color{black} m.preuss@liacs.leidenuniv.nl} ORCID: \href{https://orcid.org/0000-0003-4681-1346}{\color{black} 0000-0003-4681-1346}
}
\IEEEauthorblockA{\IEEEauthorrefmark{6}
Email: \href{evert.vn@lorentz.leidenuniv.nl}{\color{black}evert.vn@lorentz.leidenuniv.nl} ORCID: \href{https://orcid.org/0000-0003-0323-0031}{\color{black} 0000-0003-0323-0031}}
\IEEEauthorblockA{\IEEEauthorrefmark{1} Contributed equally}
}

\maketitle
\copyrightnotice

\begin{abstract}
    This paper develops and analyses a novel quantum combinatorial game: quantum checkers (codenamed \textit{Cheqqers}). 
    The concepts of superposition, entanglement, measurements and interference from quantum mechanics are integrated into the game of checkers by adding new types of legal moves.
    The addition of these new rules is done gradually by introducing several levels of `quantumness'.
    Quantum checkers provides a framework for interpolating between a known and solved classical game and a more complex quantum game, and serves as 1) a benchmark for AI players learning to play quantum games and 2) an interesting game for human players that allows them to build intuition for quantum phenomena. 
    We provide the initial analysis on the complexity of this game using random agents and a Monte Carlo tree search agent.
\end{abstract}

\begin{IEEEkeywords}
Quantum game, Quantum superposition, Quantum mechanics, Quantum computing, Quantum entanglement, Quantum interference, Combinatorial game, Game design, Monte Carlo Tree Search, Checkers, Draughts
\end{IEEEkeywords}
\section{Introduction}

Since 1925, research in quantum mechanics has resulted in numerous technological achievements, ranging from LEDs to MRI scanners. 
In the last decades our control over quantum mechanical states has become advanced enough that we can now use quantum states to do computations, resulting in the first quantum computers. 
These quantum computers could potentially solve problems that appear to be intractable on classical computers~\cite{harrow_quantum_2017}.

Although it is uncertain when useful quantum computers will be available, it is expected that they will have a large impact on society~\cite{de_wolf_potential_2017}. 
The development of quantum computers is now being steered by a small group of technical experts and policy makers, and to maximize the positive impacts of quantum computers, it is crucial to involve different members of different societal groups~\cite{roberson_talking_2021}.

It is hard for non-experts to contribute to the discussions about the future of quantum technology, because they lack intuition and understanding of quantum phenomena. 
This can be explained by the fact that quantum phenomena like superposition, measurement and entanglement are not directly encountered in daily life, in contrast to more familiar physical phenomena like gravity. 
Games can illustrate and amplify these phenomena however, and hence provide an excellent medium to build quantum intuition~\cite{mayer2019computer, piispanen_history_2023}.

In this work, we introduce quantum checkers, a novel variant of checkers that has behavior reflecting quantum phenomena. 
We hope that by playing quantum checkers, players can build intuition on quantum phenomena, empowering them to engage with the development of quantum technologies. 
Simultaneously, the game is complex enough that it provides an interesting challenge to existing AI methods; new strategies are possible with quantum moves, and interesting new representations of quantum states may arise.

Piispanen et al. have defined three dimensions in which games can be quantum~\cite{piispanenDefiningQuantumGames2025}. 
Quantum checkers is designed to have all three dimensions: the rules directly reference quantum phenomena, it can be implemented on a quantum computer, and it aims to help players build quantum intuition.

Many games have been created that are related to quantum mechanics, starting already from 1982~\cite{piispanen_history_2023}. 
This work aims to additionally fulfill another goal in games made to establish quantum intuitions. 
That goal is to make a game that is easy to learn for players without any physics background, but complicated enough to be interesting for prolonged play. 
We decided to base our game on checkers, because it is a well-known board game with the potential for complicated strategies. We codenamed the game \textit{Cheqqers}, to emphasize the \textbf{q}uantum rules of the game compared to normal checkers.

Previous work has introduced similar games based on chess~\cite{qchess, qchess2} and tic-tac-toe~\cite{qtictactoe, tiqtaqtoe}. 
Chess is a game with easy to learn rules, but leads to complicated strategies making it more difficult to master than checkers. 
Tic-tac-toe is easy to learn too, but remains too simple to allow for complicated strategies. 
We hypothesize that checkers is a good middle-ground to help people build quantum intuition in a more complicated game. 
Additionally, quantum checkers is complex enough that it makes an excellent benchmark for AI players.

In this paper we survey related work (Section~\ref{ref:related_work}), introduce the rules of quantum checkers (Section~\ref{sec:quantum_checkers}), describe our implementation (Section~\ref{sec:implementation}), and experimentally analyze the properties of quantum checkers using random agents and MCTS AI agents (Section~\ref{mcts_explanation} and 
\ref{experimental_set_up_results}).

\section{Related Work}\label{ref:related_work}

Dorbec and Mhalla propose a framework to introduce quantum-inspired moves in combinatorial games \cite{dorbec2017toward}. 
Burke et al. further research quantum combinatorial games, focusing on ``quantum flavor D'', which allows both classical and quantum moves \cite{burke2020quantum}. 
We will also use this ruleset in developing quantum checkers, i.e. classical moves are always also valid moves, and the game could hence be played fully classically too. 
This means that the number of game states in our quantum version is at least as large as the classical version.

Quantum Chess also includes superposition, entanglement, measurement and interference, and our game is heavily inspired by it; in fact, we would recommend this article to readers interested in the mathematical underpinnings of these quantum games \cite{qchess}. In other versions of quantum chess \cite{qchess2, qchess22}, all pieces exist in a superposition of two piece-type states and collapse to a classical state when touched. 

Quantum minesweeper~\cite{qminesweeper} and quantum versions of tic-tac-toe~\cite{qtictactoe, tiqtaqtoe} also serve as educational tools for teaching quantum physics. 
Quantum minesweeper involves figuring out the layout of mines in superposed classical boards, and  
quantum versions of tic-tac-toe involve placing marks in superposition, which collapse to a classical state upon measurement. 
In TiqTaqToe~\cite{tiqtaqtoe} measurements occur when all squares are occupied, and in Quantum tic-tac-toe~\cite{qtictactoe} they occur when `loops' of squares have been made.

These games share a common theme of illustrating concepts of quantum mechanics in an engaging and accessible manner. To make the discovery of quantum concepts in quantum checkers as accessible as possible, we follow the design of TiqTaqToe and incorporate different levels of quantum checkers. 
In each subsequent level, we incrementally add more quantum concepts. 
Players can then start with the lower levels and move to the higher levels as they get acquainted with the quantum concepts. 

Quantum games also serve as a playground for the development of Artificial Intelligence (AI) for quantum systems.
Though this paper will not focus on the effectiveness of games as educational tools for quantum, engagement and accessibility remain important design goals.

\section{Rules of Quantum Checkers}\label{sec:quantum_checkers}

We designed quantum checkers to be an extension of the board game checkers. 
There are many different variants of checkers, all played by two players on a grid board, but varying in the moves allowed and board size~\cite{ratrout_guide_nodate}. 
We decided to base this version of quantum checkers on English Draughts, since it is the most popular variant of checkers in the Anglophone world. We take the commonly used rules as described by the World Checkers/Draughts Federation~\cite{world_checkers_draught_federation_rules}.
In this work, we often refer to `classical checkers', and mean this version with it.
Other variants of checkers could be used to make a quantum game in a similar way, but we leave this for future work.

We designed quantum checkers to always be playable as classical checkers if the user wants to, though certain advantages may be added if quantum moves are used.
To this end, we introduce 'quantumness' in different levels. 
At level 0, there is no quantumness and the game is fully equivalent to classical checkers.
At level 1, superposition and measurements are introduced.
At level 2, another mechanic is added that allows for entangled states.
At level 3, a mechanic is introduced that allows superpositions to interfere.
These levels are constructed such, that mechanics are \emph{additive}, i.e. legal moves at lower levels are also always legal moves at higher levels.

A specific advantage to the design with different levels of quantumness is that it breaks the introduction of new mechanics into smaller steps. 
This way, a player can start with ordinary checkers to learn (or recap) the original rules. 
Then, they can move on to level 1, learning how the superposition and measurement rules work, but not yet having to deal with entanglement. 
When the player has understood this version, they can move on to higher levels of the game including more quantum concepts.

In our version detailed below, we will work with the quantum version of bits (qubits). 
The Appendix very briefly introduces these, and we refer to \cite{qchess} for a more detailed exposition. 
It is important to note here that we follow an implementation where each square is a qubit, and the qubit information tracks only whether a square is (quantumly) occupied or not. 
Piecetypes (regular vs crowned) are tracked separately. 
An implementation in which piecetype and color is included can also be made, requiring qudits with 5 states, but we have not attempted to do so.

In the rest of this Section, we will explain these sets of rules in more detail.

\subsection{Classical (no quantum) checkers}
\label{sec:classical_checkers}
We will briefly introduce the rules of English draughts, the variant of checkers we use as a basis for quantum checkers.

\subsubsection{Setup}

Checkers is a competitive game of 2 players on an $8\times 8$ board with alternating black and white tiles, of which only the 32 black tiles are used. 
The board is placed such that each player has a black square in the bottom left corner.
Each player has 12 pieces (white for one player and black for the other) placed on the first three rows of black tiles from their side.
The white player always starts, after which the players alternatingly take their turn.
In our implementation, we will consider varying board sizes smaller than $8\times 8$ to make computational requirements less stringent. 

\begin{figure}[ht]
\centering
\begin{tikzpicture}[>=latex,node distance=2em]
 \node(a){\includegraphics[width=3cm]{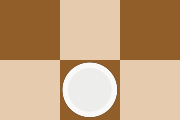}};
 \node(a){\includegraphics[width=3cm]{images/moves_white_pieces/classic_start.png}};
 \matrix[right=of a,row sep=2em] {
 \node(b){\includegraphics[width=3cm]{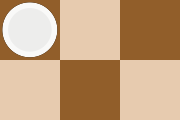}};\\
 \node(c){\includegraphics[width=3cm]{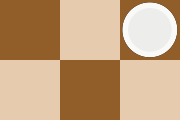}};\\ 
 };
 \draw[->] (a) -- (b);
 \draw[->] (a) -- (c);
\end{tikzpicture}
\caption{A normal move in classical checkers, where the piece moves either diagonally left or right to the first empty square. Only a subsection of a larger gameboard is shown here for illustrative purposes.}
\label{img:normal_move}
\end{figure}

\subsubsection{Movement}
When it is their turn, a player decides to make a move with one of their pieces. 
The rules for this are as follows:
\begin{itemize}
    \item Pieces can move \textbf{forwards} on diagonals to an empty square (Fig.~\ref{img:normal_move}).
    \item Pieces reaching the opposite side of the board are crowned.
    \item Crowned pieces can move forwards and backwards on diagonals.
\end{itemize}

\subsubsection{Capturing}
\begin{figure}[b]
\centering
\begin{tikzpicture}[>=latex,node distance=2em]
 \node(a){\includegraphics[width=3cm]{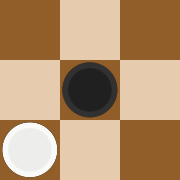}};
 \node[right=of a](b){\includegraphics[width=3cm]{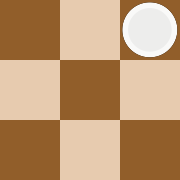}};
 \draw[->] (a) -- (b);
\end{tikzpicture}
\caption{A move where the white player takes a piece of the black player.}
\label{img:classic_capture}
\end{figure}

Opposing pieces can (or must) be captured according to these rules:
\begin{itemize}
    \item A piece can capture an opponent piece if it is adjacent on a diagonal that the piece is allowed to move in, and the square directly behind the opponent piece is empty.
    \item If one of the player's pieces can perform a capture, the player is \textit{obligated} to capture.
    \item Capturing is done by moving to the empty square behind the opponent's piece ('jumping over the piece') and removing the opponent piece (Fig.~\ref{img:classic_capture}).
    If, from this position the piece can capture another opponent piece, it must do so. If multiple captures are possible, one is chosen by the player.
    This repeats for as long as captures are possible.
\end{itemize}

\subsubsection{End of the game}
A player wins and the game ends when their opponent has no pieces left or has no legal moves on their turn. 
The game ends in a draw if more than 40 moves have been played without any captures.

\subsection{Quantum Checkers Level 1}\label{sec:superposed_checkers}

At quantumness level 1, we introduce only mechanics based on superposition and measurements. 
These rules should be understood as additional to the rules from the previous subsection, i.e. legal moves in classical checkers are still legal at this level.  

In quantum mechanics any system can be in a \emph{superposition} of multiple different states, until it is \emph{measured}, at which point it chooses one of the states it was in with some probability.
The Appendix details how this works on the level of quantum bits.
Because the act of measurement breaks down superpositions, it has a special status in quantum mechanics. 
For a more detailed introduction to quantum mechanics, we recommend \cite{quantum_atlas}. 
To integrate these phenomena into checkers we add the following rules.

\subsubsection{Movement}
\label{superposed_movement}
\begin{figure}[htb]
\centering
\begin{tikzpicture}[>=latex,node distance=2em]
 \node(a){\includegraphics[width=2cm]{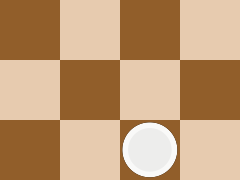}};
 \node[right=of a](b){\includegraphics[width=2cm]{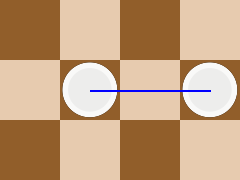}};
 \node[right=of b](c){\includegraphics[width=2cm]{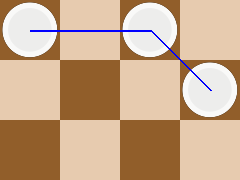}};
 \draw[->] (a) -- (b);
 \draw[->] (b) -- (c);
\end{tikzpicture}
\caption{A move where the white player moves their piece into superposition multiple times. The pieces are connected by a blue line to indicate that it is the same piece in superposition.}
\label{img:split_move}
\end{figure}

\begin{itemize}
    \item A piece can move to two squares simultaneously if both those squares are valid legal moves. 
    The piece is then split into two equal parts. 
    This move is performed by pressing the button with two arrows to the target squares.
    \item A split piece part can move as a classical piece.
    \item A split piece part can split again, as demonstrated in Fig.~\ref{img:split_move}. 
    Its probabilities are then divided, such that the resulting split pieces each have half of the original part.
\end{itemize}

\subsubsection{Measurement}
\label{superposed_measurement}
When a piece is measured (see subsection on \textit{Capturing} just below for conditions), it reverts to a classical state.
Only one of its possible positions will be true, and all other squares will be emptied. 
The probability of it being on a specific square is equal to the probability the split part on that square had as determined by the quantum mechanical rules. For more information see the Appendix.

\subsubsection{Capturing}
\label{superposed_capture}
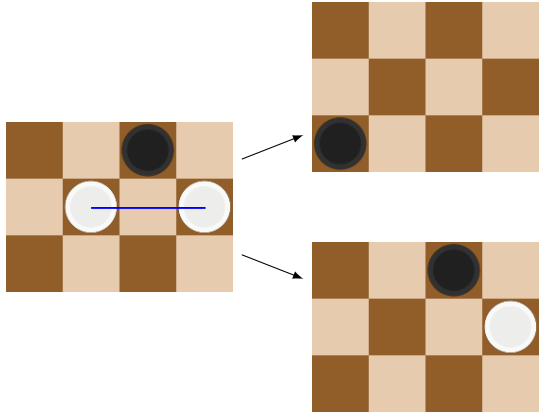
\begin{figure}[htb]
\centering
\begin{tikzpicture}[>=latex,node distance=2em]
 \node(a){\includegraphics[width=3cm]{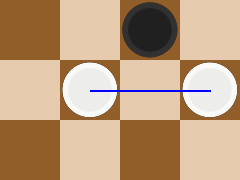}};
 \matrix[right=of a,row sep=2em] {
 \node(b){\includegraphics[width=3cm]{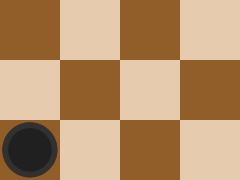}};\\
 \node(c){\includegraphics[width=3cm]{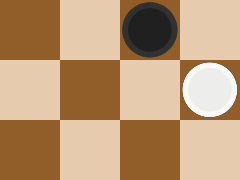}};\\ 
 }; 
 \draw[->] (a) -- (b);
 \draw[->] (a) -- (c);
\end{tikzpicture}
\caption{Two outcomes for a black piece trying to take a white piece in superposition for quantum checkers level 1.}
\label{img:split_take}
\end{figure}

Measurements are triggered in the following situations:
\begin{itemize}
    \item If an unsplit piece tries to capture a split piece, the split piece is measured.
    If the piece collapses on the square that is being captured, it is captured; otherwise, the turn counts as a pass (see Fig~\ref{img:split_take}).
    \item If a split piece tries to capture an unsplit piece, the split piece is first measured. 
    If the piece is in the place from which the capture was initiated, the capture happens normally; otherwise, the turn counts as a pass.
    \item If both pieces are split, a measurement happens on the capturing piece first. 
    If it is there, a measurement happens on the piece being captured. If it is not there, the piece being captured is not measured.
\end{itemize}

\subsection{Quantum Checkers Level 2}\label{sec:entangled_checkers}

The second level includes rules based on the quantum phenomenon of entanglement. 
These rules are additional to the rules of level 1.

At this level, split pieces are no longer automatically measured when an attempt is made to capture them by an unsplit piece.
Rather, here we encounter a special kind of conditional move that results in entanglement.
Let us consider a split piece (i.e. it is in superposition between two squares), and a move in which it is captured by an unsplit piece.
There are two scenarios:
1) The split piece will indeed be in the location where it is being taken. The split piece should then be captured and the taking piece should jump over it.
2) The split piece is not in the location where it is being taken. The taking piece should then not move.
At level 2, the result is a superposition between these two. 
That means that the piece that is capturing \textit{is now part of that superposition}: in one case it has captured, and in the other state it has not moved. 
This is an entangled state: it is not the same as both pieces being split individually, but is a correlated state in which the state of one piece depends on the other.
Note however, that in both parts of the superposition, the piece that is being taken is not on the square where it is being taken anymore, so that part of the superposition is immediately removed.

Entanglement is a striking phenomenon, because it shows that the behavior of quantum systems can not be described by any local realist theory. Furthermore, entanglement is an important resource for quantum computing. 
It has been shown that entanglement is necessary for a quantum computer to offer exponential speed-up compared to classical computing~\cite{jozsa_role_2003}.

\subsubsection{Movement}\label{entangled_movement}
\begin{itemize}
    \item An entangled piece can do a classical move or move into a superposition.
\end{itemize}

\subsubsection{Capturing}\label{entangled_capture}
\begin{figure}[htb]
\centering
\begin{tikzpicture}[>=latex,node distance=2em]
 \node(a){\includegraphics[width=3cm]{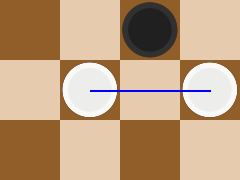}};
 \matrix[right=of a,row sep=2em] {
 \node(b){\includegraphics[width=3cm]{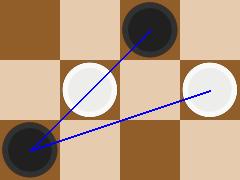}};\\
 }; 
 \draw[->] (a) -- (b);
\end{tikzpicture}
\caption{A black piece tries to take an opponent's piece in superposition. The black and white pieces become entangled as indicated by the blue line.}
\label{img:ent_take}
\end{figure}

\begin{itemize}
    \item If a classical piece tries to capture a split piece, instead of a measurement happening, both pieces become entangled (Fig.~\ref{img:ent_take}).
    \item In all other cases, an entangled piece behaves the same way as a split piece, including in regards to measurements.
\end{itemize}

\subsection{Quantum Checkers Level 3}

The third level includes rules based on the quantum phenomenon of interference. 
These rules are additional to the rules of level 1 and 2.

Apart from being probabilistic, quantum states also have a \emph{phase}. The phase itself has no influence on the outcomes of a measurement, but when different parts of the superposition with different phases combine, it does have an effect. Different parts of the superposition can then constructively or destructively interfere, causing the chance of a piece to be on a specific square to go up or down.

\begin{figure}[htb]
\centering
\begin{tikzpicture}[>=latex,node distance=2em]
 \node(a){\includegraphics[width=2cm]{images/moves_white_pieces/split_start.png}};
 \node[right=of a](b){\includegraphics[width=2cm]{images/moves_white_pieces/split.png}};
 \node[right=of b](c){\includegraphics[width=2cm]{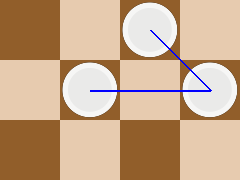}};
 \draw[->] (a) -- (b);
 \draw[->] (b) -- (c);
\end{tikzpicture}
\caption{A move where the white player moves their piece into superposition and then merges them back. The pieces are connected by a blue line to indicate that it is the same piece in superposition.}
\label{img:merge_move}
\end{figure}

Although every move in the different levels of quantum checkers causes a rotation of the phase on that piece, as explained in the Appendix, in level 1 and 2, this doesn't cause any effects. This is because the different parts of the superposition were not allowed to overlap on the same square. In this level, we introduce a new move called the \emph{merge move}, which allows us to merge two parts of a superposition to the same square under certain conditions. Due to phase rotations the merge move may not move the whole superposition away from the source squares, but instead redistributes probabilities. 
The merge move is illustrated in Fig.~\ref{img:merge_move}.

Interference is a truly quantum phenomenon and can not be simulated probabilistically. This is why in our code we have to calculate the actual quantum mechanical state, instead of being able to simply calculate the exponentially cheaper probabilities.

\subsubsection{Merge move}
\begin{itemize}
    \item If two split pieces of the same type and color, which are split from the same original piece, can move without capturing to the same square, the player can also choose to do a merge move.
    \item This move merges the two source pieces to the target square, but might leave part of the superposition on the source squares. Quantum mechanically the merge move is the inverse of the split move, for more information see the Appendix.
\end{itemize}

\section{Implementation}\label{sec:implementation}

We implemented Cheqqers in Python, using  \texttt{cirq}. 
Our source code is open source and available on GitHub~\cite{githubquantumcheckers}. 
Python was chosen because it allows the game to be playable on many different platforms. 
We use \texttt{cirq} to calculate the quantum mechanical states that are used in the game.
It also makes it possible to run quantum checkers directly on a sufficiently powerful quantum computer.

We have also implemented a web front-end for Cheqqers, so it can be played in any browser, both on smartphones and desktop computers. This web application can be run locally from our source code, or can be accessed directly on \url{https://cheqqers.com}.

\subsection{Code}

Computing quantum mechanical states is very computationally intensive (and is why quantum computers are eventually required). 
In principle, the computational cost of computing a state over $n$ connected qubits grows exponentially over $n$. 
Because of this exponential growth, exactly calculating the state of a specific superposition over 50 qubits is already too costly for an average consumer computer. 
So it is not feasible to treat the entire game state as a quantum mechanical state, as the game state can not be represented in less than 50 bits.
However, not all squares need to be treated quantum mechanically; the ones that are empty or those that are not involved in superpositions can be treated independently.

In implementing quantum checkers therefore, we instead take care to only treat those parts of the state quantum mechanically that are actually in a superposition. 
The operations leading to the state are tracked on every move, and only when the state is measured, we use \texttt{cirq} to construct a quantum circuit that is as small as possible to determine the outcome of the measurement.

In the quantum circuit, the moves are represented by quantum gates. 
We use a similar set of quantum gates as was used in Cantwell's Quantum Chess~\cite{qchess}.

We made sure to implement the different rules in a modular way, such that all three versions (classical checkers, and quantumness levels 1, 2, and 3) can be played from the same code base.

\subsection{User interface}

One of the main challenges in quantum games is to represent these phenomena visually to the players. 
We wanted to represent the split and entangled pieces, in an understandable way to make it as easy as possible for players to build intuition for these concepts. 
Although the probabilities of finding pieces on different squares can be correlated in complicated ways, we want to help understand how likely they are to find pieces on any position. To facilitate this, we display the chance of the square to be occupied on measurement in the square. This information helps the players make tactical choices when considering a superposition.

Split and merge moves act on more than two squares, so they require new interface elements for the user to select them in an intuitive way. We chose to display two arrows to the squares the piece is split to as a button for the split move. For the merge move we similarly chose to display a button with two arrows from the pieces that are merging to the merge location. Future research could investigate alternative ways of intuitively selecting these quantum moves, for example by swiping between source and target squares.

In level 3, the relative phases of the different parts of the superposition become important. 
The relative phases of a selected piece can be represented visually with an overlay (as is done in Quantum Chess~\cite{qchess}), but we have not yet implemented this.
Visual cues for quantum phases is an interesting future avenue to explore. 

\section{AI agents for quantum checkers}\label{mcts_explanation}

We implemented four different AI agents for quantum checkers: 

\begin{itemize}
    \item A random agent. This agent simply selects a randomly allowed move with uniform probability.
    \item MCTS agents. Monte Carlo Tree Search~\cite{mcts} (MCTS) is a multi-armed bandit search algorithm used for decision making, especially in board games. It has been implemented in games like Chess, Poker, Settlers of Catan, and Othello \cite{montereview}, and combined with deep reinforcement learning famously achieved expert-level play in Go \cite{silver2016mastering}. 
    We consider three MCTS agents: a \textit{low} MCTS agent with a rollout value of 200, a \textit{medium} MCTS agent with a rollout value of 400 and a \textit{high} MCTS agent with a rollout value of 800. 
    All three have an exploration parameter $c$ of $\sqrt{2}$.
\end{itemize}

These agents are useful for two reasons. 
First, they allow people to play quantum checkers and build quantum intuitions without a human opponent. 
Second, having different agents allows us to analyze the complexity and strategies of quantum checkers in an automated way, which we now turn to in the next section.

\section{Experimental setup and analysis}
\label{experimental_set_up_results}
We conducted a set of three experiments on quantum checkers, using a combination of random and MCTS agents.
This section discusses the setup and results for these experiments.
They are:
\begin{enumerate}
\item Measuring game complexity by having two random agents play on different board sizes for each level of quantumness (Section \ref{complexity}).
\item Evaluating the MCTS agent's performance by having it play against random  agents (Section \ref{mcts_performance}).
\item Calculating TrueSkill ratings by having random and MCTS agents play in a tournament (Section \ref{Trueskill}).
\end{enumerate}
The white player starts in all experiments.
Experiments are performed on the various levels of quantum checkers (described in section~\ref{sec:quantum_checkers}). 
The games are played by the four agents introduced in Section~\ref{mcts_explanation}.

\subsection{Experiment 1: Complexity}
\label{complexity} 

\begin{figure}[ht]
    \centering
    \includegraphics[width=\linewidth]{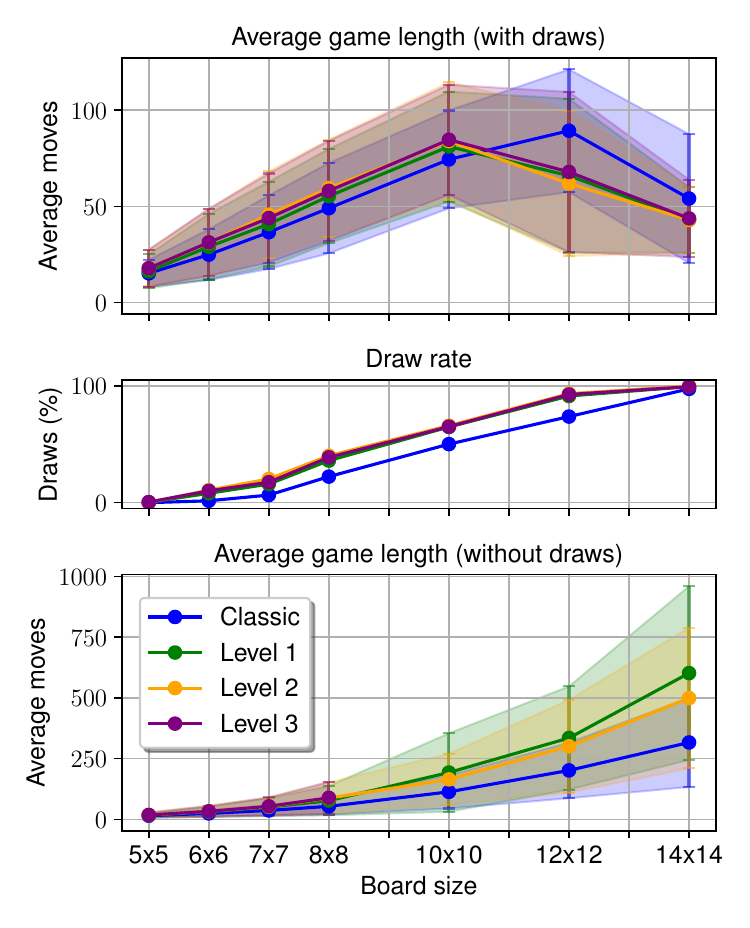}
    \caption{The average game length of 1000 games between two random players for each board size. The colored bands represent the standard deviations of the amount of moves in a game. Quantum behavior for level 1 and 2 without draws was simulated probabilistically up from size 12x12 and 10x10 respectively to reduce the computational complexity. Level 3 cannot be probabilistically simulated and is too computationally complex without draws on larger boards.}
    \label{fig:random_games}
\end{figure}

We measure the behavior of two random agents on different board sizes. 
For a direct comparison, only the first row for each player will be filled irrespective of the board size. We run a 1000 games between two random agents for each board size and quantumness level. 
The average game length with and without draws, as well as the draw rate are shown in Fig.~\ref{fig:random_games}.

The length of the game initially grows with the board size, but shrinks again with larger than $8 \times 8$ boards. The other subplots show that this decrease is due to the increase in draws because of the 40 move rule. As the board size increases it is more likely that the random agents spend 40 moves without any captures soon in the game. If we disable the 40 move draw rule, the game length keeps growing as the board size increases.

With the 40 move draw rules enabled, the quantum versions of the game have a longer average length. This is due to the fact that the game usually ends when all pieces of one player are captured. Moving pieces in superposition has a lower chance of resulting in a capture on average, because the piece might not be measured on the required location when a capture is attempted.

In the experiment, the draw chance of the quantum version of the game is higher than in classical checkers. This is also because each random move has a lower chance of leading to a capture when superpositions are present. Thus the 40 moves without a capture can be reached more easily in the quantum versions. If draws are considered undesirable, the amount of moves without a capture for a draw can simply be increased.

\subsection{Experiment 2: MCTS performance}\label{mcts_performance}
\begin{figure}[ht]
    \centering
    \includegraphics[width=\linewidth]{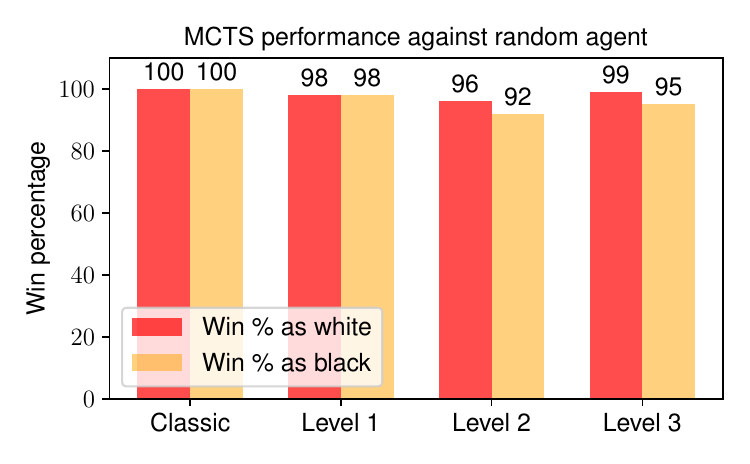}
    \caption{Win rate as percentage of MCTS agent against random agent both as the white and black player on a $5 \times 5$ board over 100 games.}
    \label{fig:mcts_performance}
\end{figure}

For this experiment we fix the MCTS agent parameters to a  budget of 800 rollouts and $c = \sqrt{2}$. Fig.~\ref{fig:mcts_performance} shows the MCTS agent's performance as white and black player against a random agent. 
The MCTS agent performs slightly better when playing as white, indicating a slight advantage for white. 

The MCTS agent wins the majority of the games, which shows that it is a feasible AI technique in quantum checkers. However, in the higher levels, it does not always win against the random agent, especially as the black player. We speculate that this is because of the higher average branching factor in the higher levels of quantumness. This causes less of the game tree to be explored with the same rollout budget. Future work could explore improving the agent by adding early termination scoring with heuristics or adding a heuristic predictor to the node selection.

\subsection{Experiment 3: TrueSkill}\label{Trueskill}
TrueSkill is a ranking system for competitive games~\cite{herbrich2006trueskill}. 
We used four agents and compared their performances using TrueSkill:

\begin{itemize}
    \item Random agent
    \item Low MCTS agent: 200 rollouts
    \item Medium MCTS agent: 400 rollouts
    \item High MCTS agent: 800 rollouts
\end{itemize}

In TrueSkill, a player's skill is represented by a Gaussian distribution and hence by two parameters: $\mu$, representing the `average skill' of a player and $\sigma$, representing the `uncertainty' in the predicted skill of the player.
Each agent starts with a default rating given by $\mu = 25$ and a confidence $\sigma = 25/3$. 
The skill level adjusts with wins or losses, and the confidence rating decreases with each game, indicating more accurate ratings.

\subsubsection{Results 5x5}
\begin{figure}
    \centering
    \includegraphics[width=\linewidth]{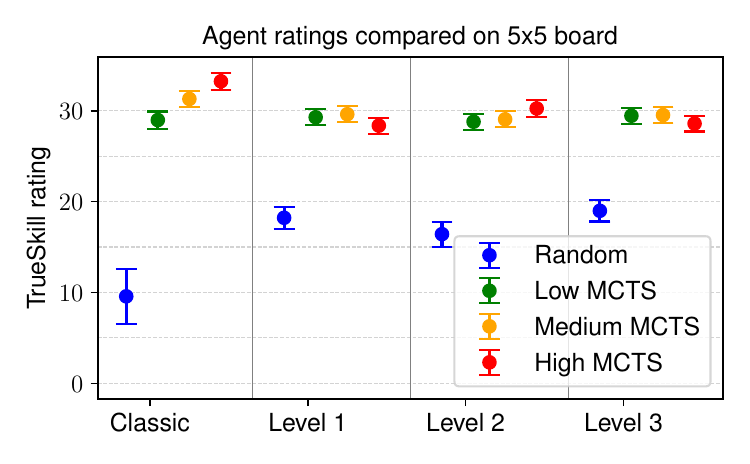}
    \caption{TrueSkill rating in each level for 150 games per agent over a total of 300 matches on a $5 \times 5$ checkerboard with one row of pieces. The error bars represent the $\sigma$ values of the TrueSkill rating.}
    \label{fig:trueskill5x5}
\end{figure}

\begin{figure}
   \centering
    \includegraphics[width=\linewidth]{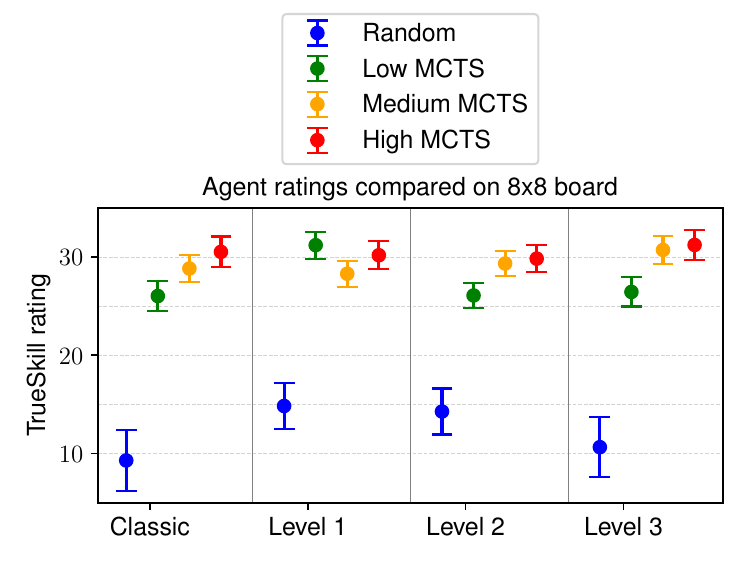}
    \caption{TrueSkill rating in each level for 48 games per agent over a total of 96 matches on a standard $8 \times 8$ checkerboard. The error bars represent the $\sigma$ values of the TrueSkill rating.}
    \label{fig:trueskill8x8}
\end{figure}

The TrueSkill ratings after a tournament (with 150 games played per agent) on the $5 \times 5$ board are shown in Fig.~\ref{fig:trueskill5x5} for each level of quantum checkers.
The figure shows that the random agent performs the worst, with the lowest rating across all levels. 
The random agent's rating increases in the higher levels. This is not due to the random agent performing better, but rather because the other agents perform comparably worse at higher levels.

All MCTS agents outperform the random agent at all levels. The MCTS agents with more rollout also perform better in classical checkers. On higher levels, the different rollout values seem to matter less for the performance of the agent. We speculate that the random outcomes of the measurements in the higher levels masks the relatively small skill difference of the agents on a small board.

\subsubsection{Results 8x8}
Fig.~\ref{fig:trueskill8x8} shows the TrueSkill ratings after a tournament on the standard checkers board (three starting rows of pieces) for each level.
The $\sigma$ value for each entry is higher due to the larger game tree of the $8 \times 8$ board, resulting in fewer games played and thus a lower confidence score for each rating.
Despite this, the same trend is observed: the random agent performs the worst, and the MCTS agents with larger rollouts budget perform the best in most cases.

This experiment shows that our MCTS agent also outperforms random agents on the standard board set up. However, at this board size, the agents do run into computational constraints. On an average consumer laptop, the high MCTS agent takes about a second to make a move. Although this is still sufficiently fast for casual play, it makes it hard to run large experiments to test its capabilities.

\section{Conclusion}\label{conclusion}
This paper presents a quantum version of checkers, introducing quantum computing concepts such as superposition, measurement, entanglement and interference.
We additionally performed a few initial experiments to build intuition about the complexity of the game and the performance of agents, but by no means performed an extensive analysis. 
The results indicate that the game is a promising benchmark for AI algorithms, allowing for a controlled and meaningful way to increase complexity by increasing quantumness.
Because of the stochastic nature, agents must deal with expectation values of moves and heuristics. 
We chose to implement a Monte Carlo Tree Search agent that simply averages over the rollouts, but future work could include more clever choices of averaging based on the piece probabilities.

The experiments suggest that new strategies may emerge in quantum checkers, meaning that the extra moves in the game bring new options beyond playing it as classical checkers. 
This demonstrates that quantum moves can be advantageous, and future work could focus on exploring these scenarios and on finding and interpreting the new strategies.

Finally, a different follow-up could instead focus on the educational value of the game and on the player experience. 

\bibliographystyle{IEEEtran}
\bibliography{IEEEabrv, ms}

\appendix
\label{sec:appendix_quantum}
Each square in quantum checkers is treated as a quantum bit, which is a quantum version of a classical bit. 
If we denote the two states of a classical bit as $|0\rangle$ and $|1\rangle$, the state of a quantum bit is given by
$|q\rangle = \alpha |0\rangle + \beta |1\rangle$. 
In this notation, $\alpha$ and $\beta$ are complex numbers and are called \emph{amplitudes}. 
They determine the probability $p_0$ and $p_1$ of measuring the qubit in state $|0\rangle$ and $|1\rangle$ respectively, through $p_0 = |\alpha|^2$ and $p_1 = |\beta|^2$ (and hence obey $|\alpha|^2 + |\beta|^2 = 1$.
Classical bits are a subset of quantum bits for which only $\alpha = 1$ and $\beta = 0$ or $\alpha = 0$ and $\beta = 1$ are allowed. 

For two qubits, the general quantum state is 
$|q_1 q_2\rangle = c_{00}|00\rangle + c_{01}|01\rangle + c_{10}|10\rangle + c_{11}|11\rangle$, and for three qubits the equivalent would be a superposition of all $8$ possible bitstrings each with their own amplitude.
The state $|q_1 q_2\rangle = \frac{1}{\sqrt{2}}|00\rangle + \frac{1}{\sqrt{2}}|11\rangle$ can not be written as the product of two independent qubits $|q_1\rangle = \alpha|0\rangle + \beta|1\rangle$ and $|q_2\rangle = \gamma|0\rangle + \delta|1\rangle$ (no such $\alpha,\beta,\gamma,\delta$ exist), and is called an entangled state for that reason. In the game, such states arise when superposed pieces try to perform captures.
After such a move, therefore, the fate of one of the pieces directly influences that of the other.

We interpret the entire board of quantum checkers as one with 64 qubits, yet we need only look at some of the $2^{64}$ possible states to play the game. 
For example, to describe a classical move from one square to another, we focus our attention only on the two squares involved. 
We call the first square the source square, and the second square the target. 
If the piece is fully on the source square, and the target is empty, we write the state as $|10\rangle$.
A classical move performs $|10\rangle \to i |01\rangle$. This moves the piece, and applies a phase rotation. These phase rotation are added to make interference possible.

The split move is implemented through the 'square-root-iSWAP' operator, exactly as in Quantum Chess~\cite{qchess}, which does the following. 
We again start with a piece fully on a source square, but will now split it to two targets as follows:
$|100\rangle \to \frac{1+i}{2}|010\rangle + \frac{1-i}{2}|001\rangle$.
When measured, we get either $|010\rangle$ with probability 0.5 or $|001\rangle$ with 0.5 probability.

The merge move is simply the inverse of the split operator. This has as a result that if two split pieces are merged together immediately after splitting, there will be no chance of them being measured on the source squares directly after. However, if they have moved since splitting, they might not merge together fully because of phase rotations from their moves.

It is too computationally expensive to  calculate the gates and superpositions over all 64 squares at the same time. 
That is also not required, unless the full game ends up in an entangled state.
Instead, the game factors into independent sets of squares that need to be simulated. 

\end{document}